\documentclass[aps,prl,twocolumn,amsmath,amssymb]{revtex4-1}

\usepackage{amssymb}
\usepackage{graphicx}
\usepackage{tabularx}
\usepackage{bm}
\usepackage{epsfig}
\usepackage[ansinew]{inputenc}
\usepackage{lineno}

\begin{document}

\title{A stretched exponential-based approach for the magnetic properties of spin glasses}

\author{L. Bufai\c{c}al}
\affiliation{Instituto de F\'{\i}sica, Universidade Federal de Goi\'{a}s, 74001-970, Goi\^{a}nia, GO, Brazil}

\date{\today}

\begin{abstract}

The spin glasses show intriguing characteristic features that are not well understood yet, as for instance its aging, rejuvenation and memory effects. Here a model based on a stretched exponential decay of its magnetization is proposed, which can describe the main magnetic features of spin glasses observed in experiments as the time-decay of thermoremament magnetization, the relaxation of zero field cooled magnetization, the ac and dc magnetization as a function of temperature and others. In principle, the here proposed model could be adapted to describe other glassy systems.

\end{abstract}

\maketitle

\section{INTRODUCTION}

The spin glass (SG) is another case in physics for which the effect of time ($t$) may bring puzzling consequences. What in the early 1960 decade seemed to be just a different class of dilute magnetic alloys exhibiting unusual magnetic susceptibility and specific heat curves, was a few latter recognized as a complex system, with some of its intriguing behavior being analogous to the mechanical properties of real glasses, showing for instance aging, rejuvenation and memory effects \cite{Mezard,Mydosh}. This disordered and frustrated system was soon stablished as a playground for both experimentalists and theorists, and the development of models and mathematical tools attempting to explain it has found application not only for SG but also in other complex systems as neural networks, protein folding and computer science \cite{Stein}. 

The two mainstream theoretical pictures used to explain the SG are the droplet-scaling model \cite{McMillan,Fisher} and the extensively investigated mean field Sherrington-Kirkpatrick model \cite{Mezard,SK} with its replica symmetry breaking derived from the Parisi's solution \cite{Parisi,Parisi2}. While analytical investigations suggest a single pair of spin-flip related states at low temperatures ($T$) as described by the first model \cite{Newman}, many computational simulations give evidence in favor of the latter with its multitude of pure states \cite{Berg}. Regarding these and the several other proposed models, and in spite of the great progress observed along these nearly five decades of investigation, as one goes deeper in these theories it feels that many of the results are poorly (if at all) connected to those obtained in laboratory. More importantly, each theory is better suited to describe a sort of SG properties as it contradicts other features. Consequently, some of the intriguing properties of SG materials are not well understood, in special those related to its dynamics.

Here an alternative approach is used to describe the magnetic properties of SG. Motivated by experimental results, a function is proposed to directly describe the systems' magnetization ($M$) after the application/removal of an external field ($H$). It is the first model that can, alone, fairly reproduce the main striking magnetic features of SG, \textit{i.e.} the thermoremanent $M$ (TRM), the zero field cooled (ZFC) $M$ (M$_{ZFC}$) and the ac and dc $M$ as a function of $T$ curves [M(T)]\cite{Mydosh}, as well as other important experiments.

\section{RESULTS AND DISCUSSION}

The model considers that if a SG system was subject to $H$ during a finite time interval $\delta t$ = $t_2 - t_1$, its $M$ at a posterior instant $t$ will be given by
\begin{equation}
M(t)=\int_{t_1}^{t_2} M_{0}e^{-b(t-t')^{n}} \,dt', \label{Eq1}
\end{equation}
where $0\leq n \leq 1$ and both $M_0$ and $b$ depend on $T$ and $H$ at $t'$:
\begin{equation}
\begin{split}
M_0 & = \left[\frac{T(t')}{T_{g}}\right]^n\frac{AH(t')}{t'-t_{g}}; \\
b & = \frac{c \cdot T(t')}{T_{g}(t'-t_{g})^n}, \\ \label{Eq2} \\
\end{split}
\end{equation} 
where $A$ is a constant dependent of the material's properties, as the constituent elements, the density of unpaired moments etc. Although a more profound understanding of the implications of this proposed model is desired before any assumption concerning its physical origin, one may speculate, roughly speaking, that the decay expressed in Eq. \ref{Eq1} could be related to the search for lower-energy states through the systems' rugged energy landscape, where the $t'-t_g$ term plays the role of aging, \textit{i.e.} the system is continuously evolving after the transition temperature $T_g$ was achieved at instant $t_g$. The $c$ parameter is expected to depend on $H$, since changing it leads the system to a different position in the energy landscape, thus affecting its relaxation. But as the main part of this study is dedicated to situations in which $H$ is constant, the discussion of such variable will be postponed to section 2.3. The $n$ parameter, together with $c$ and $T/T_g$, determine the systems' glassiness, \textit{i.e.} how slow $M$ will decay. 

At a first glance, it may look that this model keeps close resemblance with the stretched exponential decay multiplied by a power law of $t$ \cite{Ocio}
\begin{equation}
M(t)=C\left(\frac{t}{t_{p}}\right)^{-\alpha} \cdot e^{-(t/t_{p})^{n'}}, \label{Eq3}
\end{equation}
and to its variants that are usually adopted to fit TRM and M$_{ZFC}$ curves \cite{Chamberlin,Nordblad}. However, there are some remarkable differences between the here proposed model and previous ones, the most significant one being the fact that here the magnetization is the outcome of an integration along the interval during which $H$ was applied. Moreover, those previous models are only suitable to fit the TRM and M$_{ZFC}$ curves whereas the here described one is proposed to be more general, enabling the description of other experimental results, as will be discussed.

\subsection{Thermoremanent Magnetization}

Beginning with the TRM experiment, a typical TRM curve is carried after cool the system from above $T_g$ down to a measuring $T$ ($T_m$) in the presence of $H$. After keeping the system at this condition for a waiting time $t_w$, $H$ is removed (at $t = 0$) and the remanent $M$ is recorded as a function of $t$ (for a visual description of this protocol see the Supplementary Material - SM \cite{SM}). Fig. \ref{Fig_TRM}(a) shows the curve calculated at $T_m$ = 0.8$T_g$ with $n$ = 0.5 (a value within the range typically found in the fittings of TRM with the stretched exponential Eq. \ref{Eq3}), $H = A = c = 1$ (arb. units) and $t_w$ = 100 s, obtained after cool the system in a constant $T$ sweep rate $|dT/dt|$ = 0.002 $T_g$/s. It may be noticed that all parameters are given in arbitrary units with the exception of $t$, expressed in seconds (s). This is because $t$ is particularly important here in the study of the dynamics of SG, and its description in s unit facilitates the comparison of the results obtained from the model with those referenced from the experiments. The resulting curve  shown in Fig. \ref{Fig_TRM}(a) is very similar to those observed experimentally \cite{Chamberlin,Nordblad}.

\begin{figure*}
\begin{center}
\includegraphics[width=1.0 \textwidth]{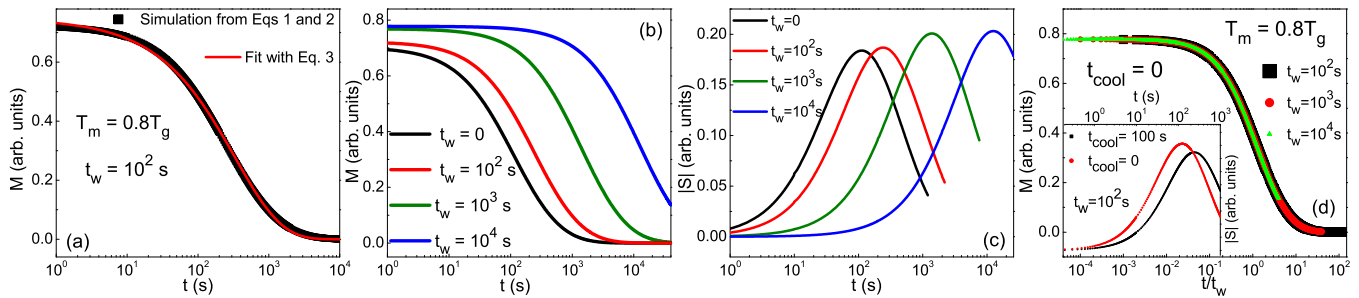}
\end{center}
\caption{(a) TRM curve calculated at $T_m$ = 0.8$T_g$ with $H = A = c = 1$ (arb. units) and $t_w$ = 100 s. The red solid line represents the best fit with Eq. \ref{Eq3}. (b) Comparison of TRM curves calculated for different $t_w$. (c) The modulus of the relaxation rate $|S|$ for the TRM curves with different $t_w$. (d) $t_{cool}$ = 0 $TRM$ curves calculated with different $t_w$, plotted as a function of $t/t_w$. The inset compares the $|S|$ for $t_w$ = 100 s TRM curves calculated with $t_{cool}$ = 100 s and with $t_{cool}$ = 0.}
\label{Fig_TRM}
\end{figure*}

For a quantitative comparison between the here proposed model and the one largely used to fit experimental TRM curves, the solid line in Fig. \ref{Fig_TRM}(a) shows a reasonably fit of Eq. \ref{Eq3} with the theoretical curve obtained from Eqs. \ref{Eq1} and \ref{Eq2}, yielding $t_p$ $\simeq$ 260 s, $n'$ $\simeq$ 0.6, these values being within the range usually found for canonical SG \cite{Ocio}. This clearly demonstrates that the proposed model is suitable to describe typical experimental TRM curves of SG materials. The fitting is not so good for small $t$, as was already observed experimentally at the early stages of investigation of SG systems, which motivated the search for alternative equations \cite{Ocio,Chamberlin,Nordblad}. It is important to note the tendency toward zero in $M$, contrasting to the experimental results showing that usually the system reach a finite magnetization value at large $t$ \cite{Chamberlin,Nordblad}. It is thus possible that, in practice, for real SG materials a fraction of the spins gets pinned toward the $H$ direction after its removal, while the other part relax. This could be easily adjusted here with the addition of a constant term.

Fig. \ref{Fig_TRM}(b) compares TRM curves calculated for different $t_w$, where a clear $t_w$-dependence is observed. This is better visualized in Fig.\ref{Fig_TRM}(c) where the modulus of the relaxation rate, $S = (1/H)(dM/dln t)$, is computed. As can be seen, a maxima in $|S|$ occurs at $t$ close to $t_w$, again reproducing the experiments \cite{Nordblad}. Such maxima is present even for $t_w$ = 0, which is due to the finite $t$ interval taken to cool the system from $T_g$ to $T_m$ ($t_{cool}$) \cite{Rodriguez,Zotev,Orbach}. As $t_w$ increases, the relative influence of $t_{cool}$ diminishes and the maxima in $S$ gets closer to $t = t_w$. If one considers the situation in which the system is immediately cooled from above $T_g$ to $T_m$ (\textit{i.e.} assuming an unrealistic $|dT/dt| = \infty$) then $t_{cool} = 0$ and the peak in $S$ will shift to the left as shown in the inset of Fig. \ref{Fig_TRM}(d). Interestingly, all TRM curves calculated for $t_{cool}$ = 0 with different $t_w$, plotted as a function of $t/t_w$, coincide [Fig. \ref{Fig_TRM}(d)], in agreement with the tendency toward full aging experimentally found \cite{Rodriguez}.

\begin{figure}
\begin{center}
\includegraphics[width=0.45 \textwidth]{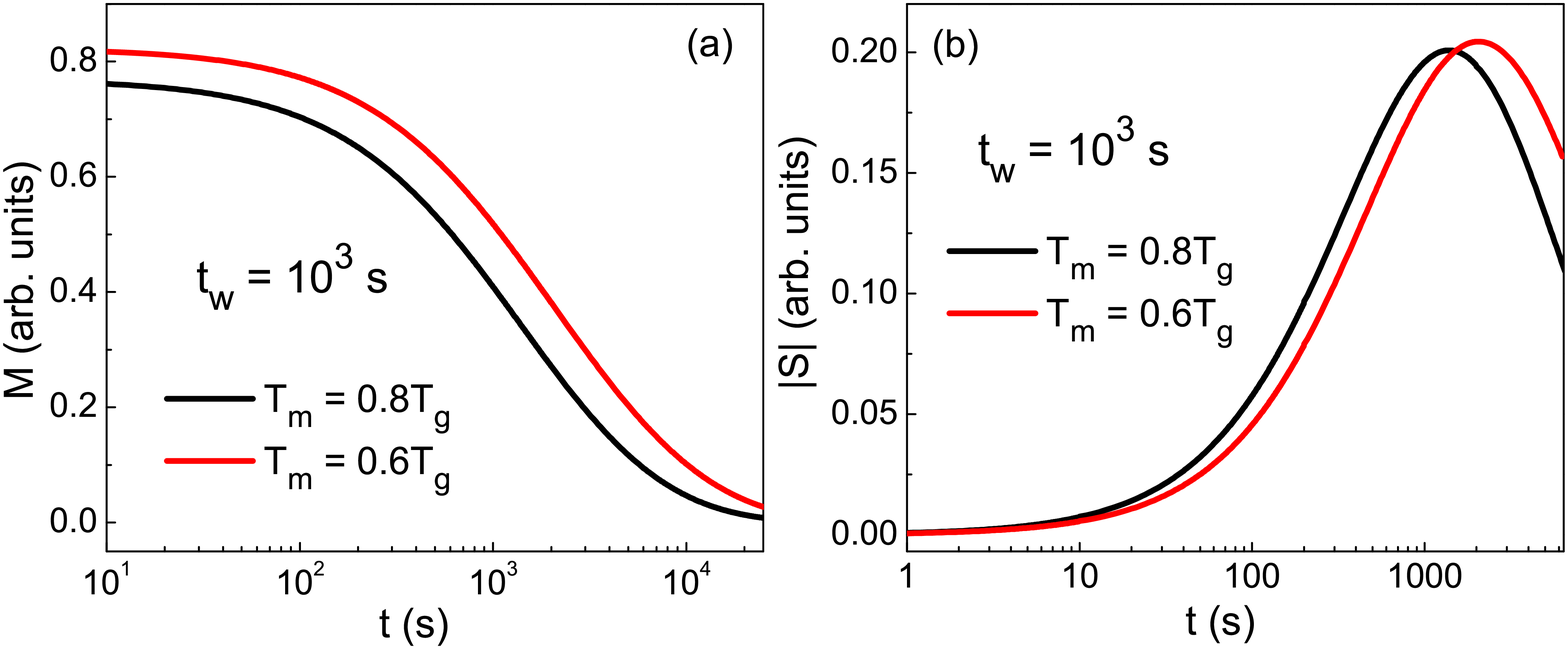}
\end{center}
\caption{(a) Comparison between TRM curves obtained with $t_w$ = 10$^3$ s and the same $H, a, c$, but distinct $T_m$. (b) The $|S|$ for these curves.}
\label{Fig_TRM_T}
\end{figure}

The model can faithfully predict the effect of thermal energy on the TRM curves. Fig. \ref{Fig_TRM_T}(a) compares the $t_w$ = 10$^3$ s TRM curves obtained with $T$ = 0.8$T_g$ and 0.6$T_g$, where it is observed the increase in $M$ for the later, while Fig. \ref{Fig_TRM_T}(b) shows its expected $|S|$ shift to larger $t$ resulting from the fact that the spins get $more$ $freezed$ with decreasing $T$, turning the decay slower. In spite of the resemblance of Fig. \ref{Fig_TRM_T}(a) with that of the great majority of SG materials \cite{Chamberlin,Nordblad}, the $T$-dependence of $M_0$ expressed in Eq. \ref{Eq2} is not expected to be universal, in the sense that there were also found materials for which the magnitude of $M$ decreases with $T$ \cite{Ocio}. One can choose other $M_{0}(T)$ functions leading to different trends for the magnitude of $M$ as $T$ changes without greatly affecting the main SG features (see SM \cite{SM}). 

\subsection{Zero Field Cooled Magnetization}

\begin{figure}
\begin{center}
\includegraphics[width=0.45 \textwidth]{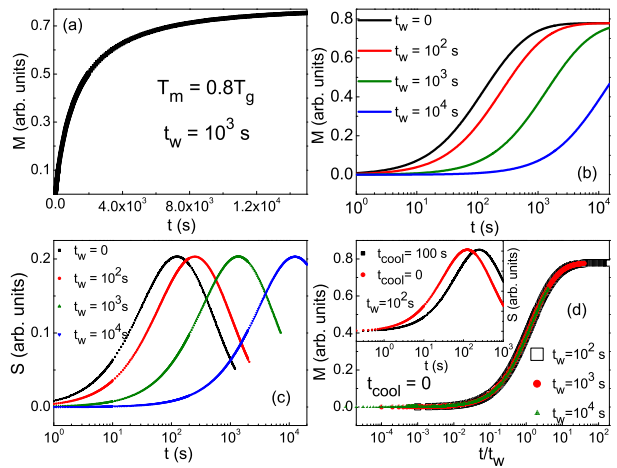}
\end{center}
\caption{(a)M$_{ZFC}$ curve calculated for $T_m$ = 0.8$T_g$, $n$ = 0.5, $H = A = c = 1$ (arb.units) and $t_w$ = 10$^3$ s. (b) Log-linear plots of the curves obtained with different $t_w$. (c) The relaxation rates, $S$, of the curves with different $t_w$. (d) M$_{ZFC}$ curves calculated for $t_{cool}$ = 0 and different $t_w$, plotted as a function of $t/t_w$. The inset compares the $S$ of $t_w$ = 100 s M$_{ZFC}$ curves calculated with $t_{cool}$ = 100 s and $t_{cool}$ = 0.}
\label{Fig_Mzfc}
\end{figure}

Besides the TRM experiments, the here proposed model can also reproduce the M$_{ZFC}$ curves, which are obtained after ZFC the system down to $T_m$ $<$ $T_g$, keep it on this condition for $t_w$, then apply a small $H$ (at $t$ = 0) and start to capture $M$ as a function of $t$ (see SM \cite{SM}). Fig. \ref{Fig_Mzfc}(a) shows the curve calculated for $t_w$ = 10$^3$ s at $T_m$ = 0.8$T_g$ and using the same parameters chosed to produce the TRM curves, \textit{i.e.} $n$ = 0.5, $H = A = c = 1$ (arb. units), resulting in a fair agreement with the typical experimental curves reported for SG system \cite{Lundgren}. From a log-linear plot of the curves obtained with different $t_w$, Fig. \ref{Fig_Mzfc}(b), one can see the expected $t_w$-dependency observed experimentally \cite{Granberg}. Fig. \ref{Fig_Mzfc}(c) displays the $S$ resulting from these M$_{ZFC}$ curves. As for TRM, the maxima in $S$ for M$_{ZFC}$ occurs at $t$ larger than (but close to) $t_w$, precisely the same behavior as that of experimental curves \cite{Granberg}. Here, although there is no magnetization during cooling since it occurs at zero $H$, $t_{cool}$ still plays its part because according to Eq. \ref{Eq2} the system starts to age already after the system passes through $T_g$ (at $t_g$). As $t_w$ increases, the relative effect of $t_{cool}$ decreases in comparison to $t_w$, and the maximum in $S$ gets closer to $t = t_w$. As in the case of TRM curves, if we assume $t_{cool}$ = 0 in the M$_{ZFC}$ protocol the plot of $M$ as a function of $t/t_w$ will indicate a tendency toward full aging, Fig. \ref{Fig_Mzfc}(d). By comparing Figs. \ref{Fig_TRM}(c) and \ref{Fig_Mzfc}(c) quantitatively it can be noticed that, as observed experimentally, the relaxation rates of TRM and M$_{ZFC}$ have nearly the same absolute values, indicating a similar aging process for both \cite{Sandlund}.

\begin{figure}
\begin{center}
\includegraphics[width=0.45 \textwidth]{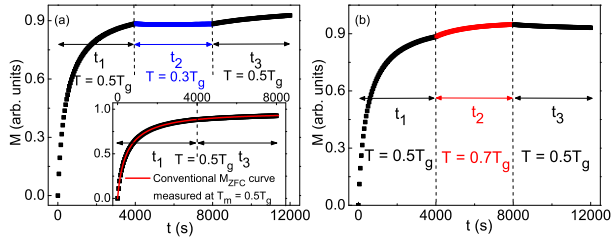}
\end{center}
\caption{(a) M$_{ZFC}$ curve calculated at $T_m$ = 0.5$T_g$ with a temporary cooling of  $\Delta T$ = 0.2$T_g$. The inset shows the curve resulting when the data at $T_m$-$\Delta T$ is removed, evidencing the memory effect. (b) M$_{ZFC}$ curve calculated at $T_m$ = 0.5$T_g$ with a temporary heating of  $\Delta T$ = 0.2$T_g$, where no memory effects appear.}
\label{Fig_Tcycle}
\end{figure}

Another strategy developed to investigate the low $T$ dynamics of SG systems is the $T$ cycling below $T_g$. Fig. \ref{Fig_Tcycle}(a) shows the curve resulting from a protocol firstly proposed to investigate memory effects in assembly of magnetic nanoparticles \cite{Sun,Khan}, in which the system is ZFC down to $T_m$ $<$ $T_g$, then a small $H$ is applied (at $t$ = 0) and the magnetic relaxation starts to be captured. After the lapse of a period $t_1$, however, the system is further cooled to a lower $T$ = $T_m$ - $\Delta T$, and kept at this condition for a period $t_2$. After the lapse of $t_2$ the system is heated back to $T_m$ and the magnetization is recorded for a period $t_3$ \cite{SM}. 

The curve in Fig. \ref{Fig_Tcycle}(a) was produced with $T_m$ = 0.5$T_g$, $\Delta T$ = 0.2$T_g$, $t_1 = t_2 = t_3 = 4000$ s and the same parameter values as those used to calculate the conventional TRM and M$_{ZFC}$ curves described above. At $t_1$ the curve is similar to those of Fig. \ref{Fig_Mzfc}, with an initial jump in the magnetization when $H$ is turned on, followed by a slow relaxation. During the temporary cooling at $t_2$, the relaxation becomes very weak, which can be inferred from the $T$-dependencies of Eqs. \ref{Eq1} and \ref{Eq2}. When the system returns to $T_m$ in $t_3$ the magnetization comes back to the level it reached before the $T$ cycling. The inset shows the curve resulting when the $t_2$ interval is removed. It makes clear the fact that during the temporary cooling the relaxation is almost halted, and the memory effect is manifested in $t_3$ when the system returns to $T_m$ and the relaxation is resumed, thus mimicking the experimental curves with precision \cite{Sun,Khan}. Conversely, for a positive $T$ cycling [Fig. \ref{Fig_Tcycle}(b)] the relaxation is hasted in $t_2$, and when the system is cooled back to $T_m$ the magnetization does not restore to the level reached before the temporary heating, also in agreement with experimental observations \cite{Sun,Khan}. These results indicate that the here proposed model may be also suitable for magnetic nanoparticles. 

The model has failed, however, to reproduce the memory and rejuvenation effects for the case of M$_{ZFC}$ experiments in which $T$ is cycled before the application of $H$ \cite{Nordblad2}, as well as the chaotic effect observed in the memory dip experiments where the ZFC process is halted prior to the measurement of M(T) \cite{Mathieu,Jonason}. It could not predict either the memory and rejuvenation effects in TRM experiments where $T$ is cycled during the measurement, because in this case $T$ is changed after the $H$ cutoff \cite{Sun,Khan}. For this last case, such contrast to the experiments suggests that the internal field may play an important role on the relaxation, and the here proposed model should be adjusted in order to take this into account. For instance, a natural attempt could be the replacement of $T(t')$ by $T(t)$ in Eqs. \ref{Eq1} and \ref{Eq2} since one may expect that, even in the absence of $H$, when $T$ is changed the energy landscape is altered and the decay will be affected (see SM \cite{SM}). This would lead to $T$ cycled TRM curves closer to the experimental ones, but would not reproduce the memory dip experiments. 

\subsection{Magnetization as a function of temperature}

\begin{figure}
\begin{center}
\includegraphics[width=0.45 \textwidth]{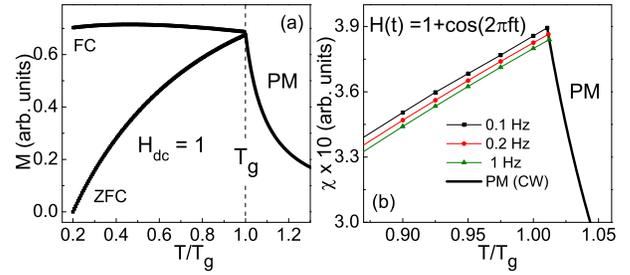}
\end{center}
\caption{(a) ZFC and FC $dc$ M(T) curves calculated with $H$ = 1. (b) ac $\chi(T)$ curves calculated for $f$ = 0.1, 0.2 and 1 Hz.}
\label{Fig_MxT}
\end{figure}

Finally, Eqs. \ref{Eq1} and \ref{Eq2} can also predict the behavior of SG systems in ac and dc M(T) experiments. Fig. \ref{Fig_MxT}(a) shows the dc ZFC and FC curves calculated for $n$ = 0.5, $H = A = c = 1$ (arb. units) and $|dT/dt|$ = 0.001 $T_g$/s. Despite the well known deviation from the Curie-Weiss (CW) behavior for the paramagnetic (PM) region of SG systems \cite{Morgownik}, for simplicity it was chosen here a CW curve for the $T > T_g$ region, which was adjusted to coincide with the $T < T_g$ ZFC and FC curves at $T_g$. The ZFC curve shows a sharp cusp while the FC one shows a plateau-like behavior, being these striking features of SG systems \cite{Nagata}. It is important to notice that the here proposed model does not predict the PM-SG transition, since it is only concerned with the SG state, $T < T_g$. The cusp-like behavior observed in Fig. \ref{Fig_MxT} results from the fact the SG curves were calculated up to $T_g$ and joined to the PM ones that were calculated only down to this critical $T$. Concerning the fact that the experimental ZFC peaks are usually sharper than that of Fig. \ref{Fig_MxT}(a) while the FC ones usually show a small bump close to $T_g$, it must be stressed that the physics for $T$ very close to $T_g$, where a divergent behavior is expected, is neither under consideration here.

According to the here proposed model, the ZFC curve depends on the cooling/heating $T$ rates (see SM \cite{SM}), as expected for an off-equilibrium condition \cite{Mydosh}. Contrastingly, the FC is nearly invariant under changes in $|dT/dt|$ and this may be the reason why it is widely believed that the FC is roughly an equilibrium situation \cite{Malozemoff, Chamberlin2,Matsui}. However, it is in fact a metastable configuration \cite{Beckman}, which can be fairly captured by the here proposed model. According to the model, if the cooling is halted for a finite $t$ interval below $T_g$ for instance, the FC magnetization will change \cite{SM}, as already observed experimentally \cite{Pal}.

Fig. \ref{Fig_MxT}(b) shows ac susceptibility curves for some selected frequencies ($f$), obtained considering an oscillating field of the form $H(t) = H_{dc} + hcos(2\pi f t)$, where $h$ is the ac field amplitude. All curves were calculated in the heating mode with $n$ = 0.5, $h = H_{dc} = A = 1$, and each point was recorded after one field cycle. The $f$ were chosen slow enough so that one can assume a nearly linear response of $M$ in relation to $H$ and use the approximation $\chi = M/H$. The stretched exponential term in Eq. \ref{Eq1} is expected to depend on $H$, thus for Fig. \ref{Fig_MxT}(b) it was used $c = |H(t)|$, but very similar curves are observed for a constant $c$ (see SM \cite{SM}). The PM curve was calculated with the same slope of that used for the dc field shown in Fig. \ref{Fig_MxT}(a), and adjusted to coincide with the $f$ = 0.01 Hz curve at $T_g$, assumed here as a nearly static situation. The resulting curves are clearly $f$-dependent, showing a tendency of decrease in magnitude with increasing $f$. Defining the freezing $T$ ($T_f$) as the point where each curve intercepts the PM curve, one can observe the expected shift of $T_f$ toward higher $T$ as $f$ increases. The relative shift $\delta T_f$ = $\Delta T_f/T_f(\Delta log f)$ \cite{Mulder} can be computed, yielding in this case a $\delta T_f \simeq$ 0.003 within the range experimentally found for canonical SG \cite{Mydosh}. Though, care must be taken with this result since it depends on the choice of the PM curve, which is known to deviate from CW behavior for SG systems \cite{Morgownik}. Moreover, it may be also related to the underline physics around $T_g$ (not considered here), so that the $T_f$ values may be related to the systems' behavior at both above and below $T_g$.

\section{CONCLUSIONS}

In summary, the model here proposed, based on a stretched exponential decay of the magnetization after the application of $H$ for an infinitesimal $t$, can describe the striking features of TRM, M$_{ZFC}$, ac and dc ZFC-FC M(T) curves and some of the memory experiments. It does not answer all the questions, thus it must be regarded as an approximate model. Nevertheless, the fact that it can reproduce several of the main SG features is remarkable, and its thorough investigation may give important insights into its physical origin, resulting in a better understanding of the microscopic mechanism behind the glassy behavior. In principle it could be also applied to other complex systems after a suitable adjustment of the parameters.
 
\section{ACKNOWLEDGMENTS}
This work was supported by Conselho Nacional de Desenvolvimento Cient\'{i}fico e Tecnol\'{o}gico (CNPq), Funda\c{c}\~{a}o de Amparo \`{a} Pesquisa do Estado de Goi\'{a}s (FAPEG) and Coordena\c{c}\~{a}o de Aperfei\c{c}oamento de Pessoal de N\'{i}vel Superior (CAPES). The author thanks Wesley B. Cardoso for the computational help.

\end{document}


\title{Supplementary Material: ``A stretched exponential-based approach for the magnetic properties of spin glasses"}

\author{L. Bufai\c{c}al}
\affiliation{Instituto de F\'{\i}sica, Universidade Federal de Goi\'{a}s, 74001-970, Goi\^{a}nia, GO, Brazil}

\maketitle
\beginsupplement

\section{Methods}

The curves displayed in this article were calculated using Maple 17 software (Maplesoft$^{TM}$, Japan), with the exception of the fitting with Eq. 3 shown in Fig. 1(a) of main text and Figs. \ref{Fig_S1}, \ref{Fig_S3}, \ref{Fig_S5}(a), \ref{Fig_S6}(a) and \ref{Fig_S7}(a) showing the protocols adopted to mimic each experiment described in text, which were performed on Origin 8.5 software (OriginLab Corporation, USA).

\section{Thermoremanent Magnetization}

\begin{figure}[h]
\begin{center}
\includegraphics[width=0.5 \textwidth]{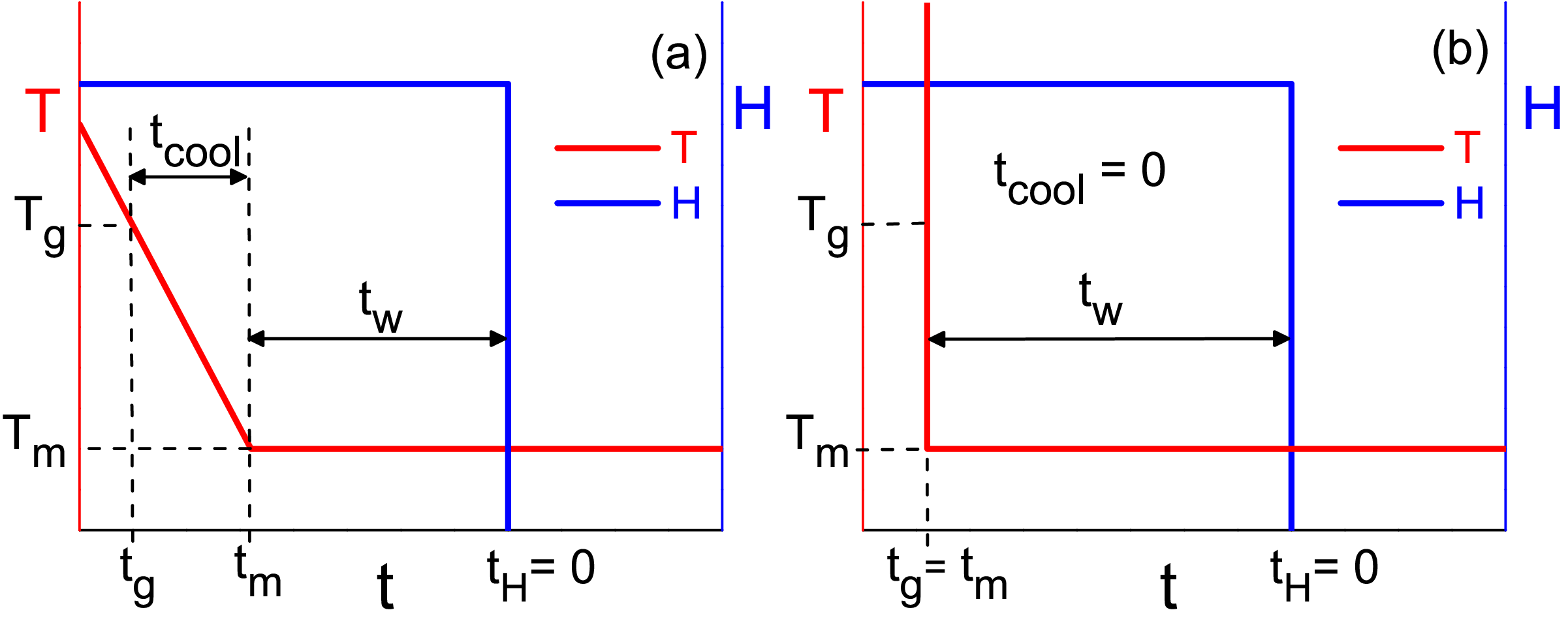}
\end{center}
\caption{The protocols adopted to produce TRM curves for the cases (a) in which $T$ decreases at a constant finite sweep rate and (b) in which $T$ is immediately quenched to $T_m$ ($t_{cool}$ = 0).}
\label{Fig_S1}
\end{figure}

Fig. \ref{Fig_S1}(a) shows the protocol used to simulate the thermoremanent magnetization (TRM) curves. The system is cooled in a constant temperature ($T$) sweep rate ($|dT/dt|$) from above the spin glass (SG) $T$ ($T_g$) down to a measure $T$ ($T_m$) in the presence of an external magnetic field ($H$). It is kept in this condition for a waiting time $t_w$, then $H$ is removed (at $t_H$ = 0) and the remanent magnetization ($M$) is recorded as a function of time ($t$). As can be noticed from Fig. \ref{Fig_S1}(a), the experimentally usual situation in which $|dT/dt|$ changes in the vicinity to achieve $T_m$ was not considered here, as well as the interval taken for the system to reach $H$ = 0 since this interval is usually very small in comparison to the measurement time, and its influence on the resulting $M$ is thus negligible.

Each TRM curve shown here results from the integration of Eq. 1 of main text along the whole $t$ interval at which $H$ was applied. Thus, using Eq. 2 on Eq. 1 one have:
\begin{equation}
M_{TRM}(t)= \int_{t_g}^{t_H = 0} \left[\dfrac{T(t')}{T_{g}}\right]^n\frac{AH}{t'-t_{g}}e^{-\frac{cT(t')}{T_g} \left( \frac{t-t'}{t'-t_g} \right)^{n}}  \,dt', \label{Eq_S1}
\end{equation}
which in this case can be divided in two integrals, one for the cooling process and other for the $T = T_m$ interval:
\begin{equation}
M_{TRM}(t)= \int_{t_g}^{t_m} \frac{AH\left[ T_g-|dT/dt|(t'-t_g)\right]^n}{T_g^n(t'-t_g)}e^{-\frac{c \left[ T_g-|dT/dt|(t'-t_g)\right]}{T_g} \left( \frac{t-t'}{t'-t_g} \right)^{n}}  \,dt' + \int_{t_m}^{t_H = 0} \left( \frac{T_m}{T_g} \right)^{n}\frac{AH}{t'-t_g}e^{-\frac{cT_m}{T_g} \left( \frac{t-t'}{t'-t_g} \right)^{n}}  \,dt'. \label{Eq_S2}
\end{equation}
It is clear from Eq. \ref{Eq_S2} that, although the weight of the first integral is usually smaller than the second one, the $t$ interval taken to cool the system from $T_g$ to $T_m$ ($t_{cool}$) plays its part in the resulting TRM curve. Fig. \ref{Fig_S1}(b) shows the protocol for the situation in which $T$ is immediately quenched to $T_g$ ($t_{cool}$ = 0), resulting in the TRM curves displayed in Fig. 1(d) of main text. In this case, $t_m = t_g$ and the first integral of Eq. \ref{Eq_S2} vanishes.

\begin{figure}[h]
\begin{center}
\includegraphics[width=0.7 \textwidth]{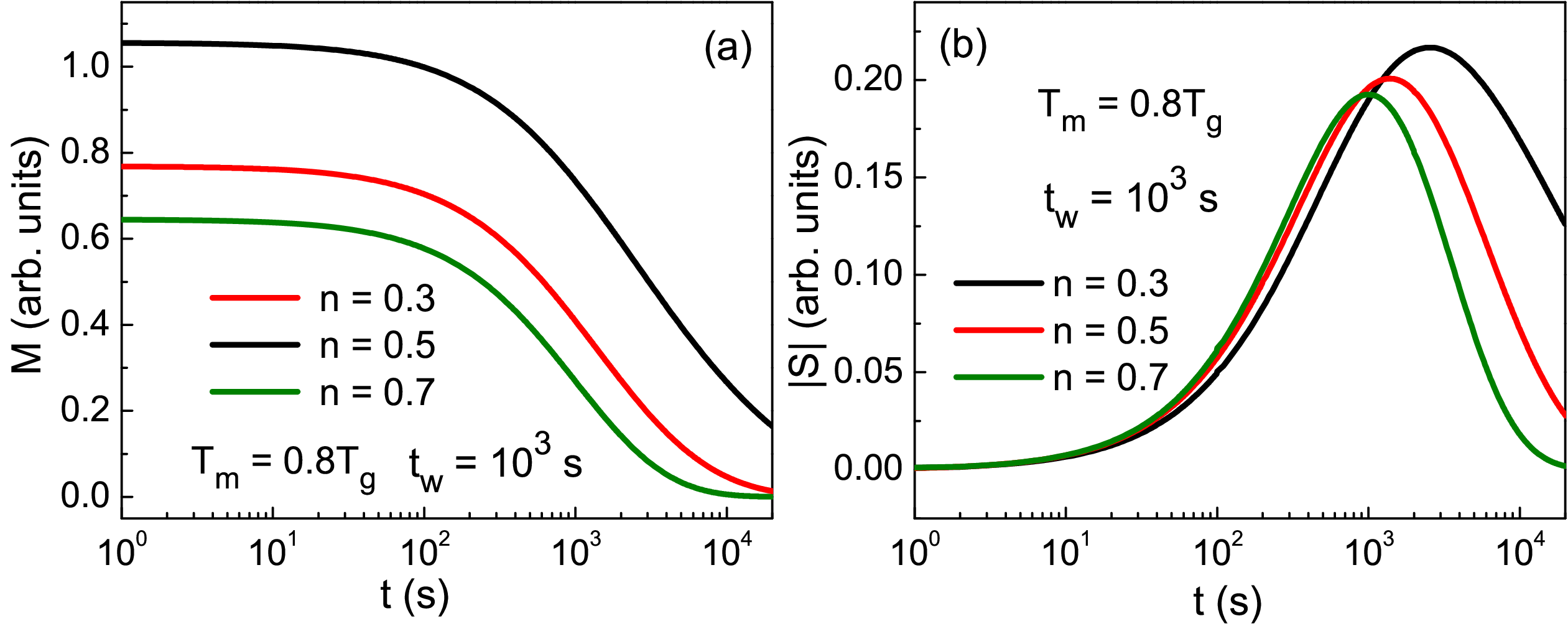}
\end{center}
\caption{(a) TRM curves calculated with $A = H = c = 1$, $T_m$ = 0.8$T_g$, $|dT/dt|$ = 0.002$T_g$, $t_w$ = 10$^{3}$ s and different $n$ values. (b) $|S|$ for these TRM curves.}
\label{Fig_S2}
\end{figure}

According to Eqs. 1 and 2 of main text, the $M$-decay will depend on $T$, $H$ and $n$. A fundamental difference from this model to other ones is that now $n$ is separated from $T$ and $H$ in the stretched exponential term. In this sense, $n$ can be understood as a characteristic of the material under study, giving a measure of its glassiness. Fig. \ref{Fig_S2}(a) shows TRM curves calculated for $A = H = c = 1$, $T_m$ = 0.8$T_g$, $|dT/dt|$ = 0.002$T_g$, $t_w$ = 10$^{3}$ s and different $n$ values. As can be noticed from Fig. \ref{Fig_S2}(b) the relaxation becomes slower as $n$ decreases.

\section{Zero Field Cooled Magnetization}

\begin{figure}[h]
\begin{center}
\includegraphics[width=0.5 \textwidth]{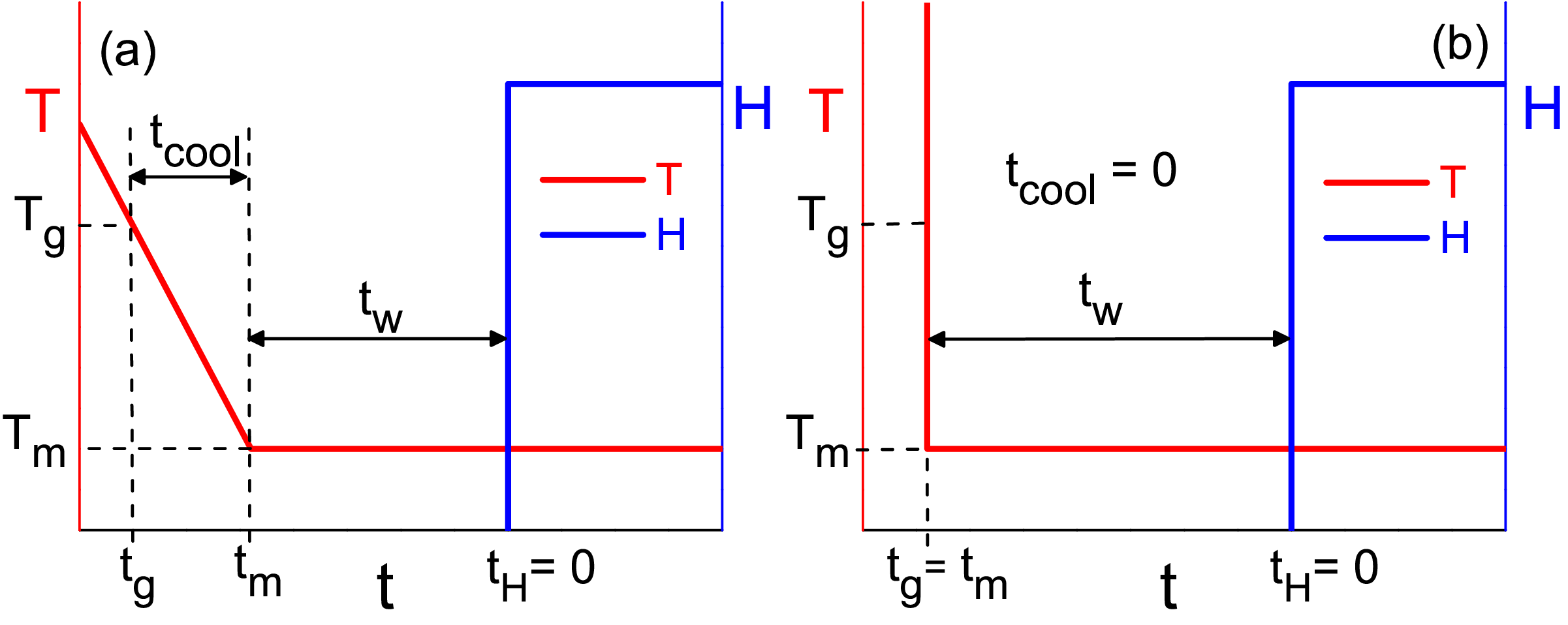}
\end{center}
\caption{The protocols adopted to produce M$_{ZFC}$ curves for the cases (a) in which $T$ decreases at a constant finite $|dT/dt|$ and (b) in which $t_{cool}$ = 0.}
\label{Fig_S3}
\end{figure}

To produce the zero field cooled (ZFC) $M$ curves (M$_{ZFC}$), the system is cooled in zero $H$ down to $T_m$ in constant $|dT/dt|$, it is kept in this condition for a $t_w$ interval, then $H$ is applied (at $t_H = 0$) and $M$ starts to be captured as a function of $t$, as shown in Fig. \ref{Fig_S3}(a). For this case, $M$ will be given by
\begin{equation}
M_{ZFC}(t)= \int_{t_H = 0}^{t} \left( \frac{T_m}{T_g} \right)^{n}\frac{AH}{t'-t_g}e^{-\frac{cT_m}{T_g} \left( \frac{t-t'}{t'-t_g} \right)^{n}}  \,dt'. \label{Eq_S3}
\end{equation}
Despite the fact the system is cooled in zero $H$, $t_{cool}$ still affects the relaxation due to the ($t'$ - $t_g$) term. As faster is $|dT/dt|$ during cooling, smaller will be the influence of $t_{cool}$ on the relaxation. Fig. \ref{Fig_S3}(b) shows the protocol used to produce the idealized $t_{cool}$ = 0 M$_{ZFC}$ curves displayed in Fig. 2(d) of main text.

\begin{figure}
\begin{center}
\includegraphics[width=0.6 \textwidth]{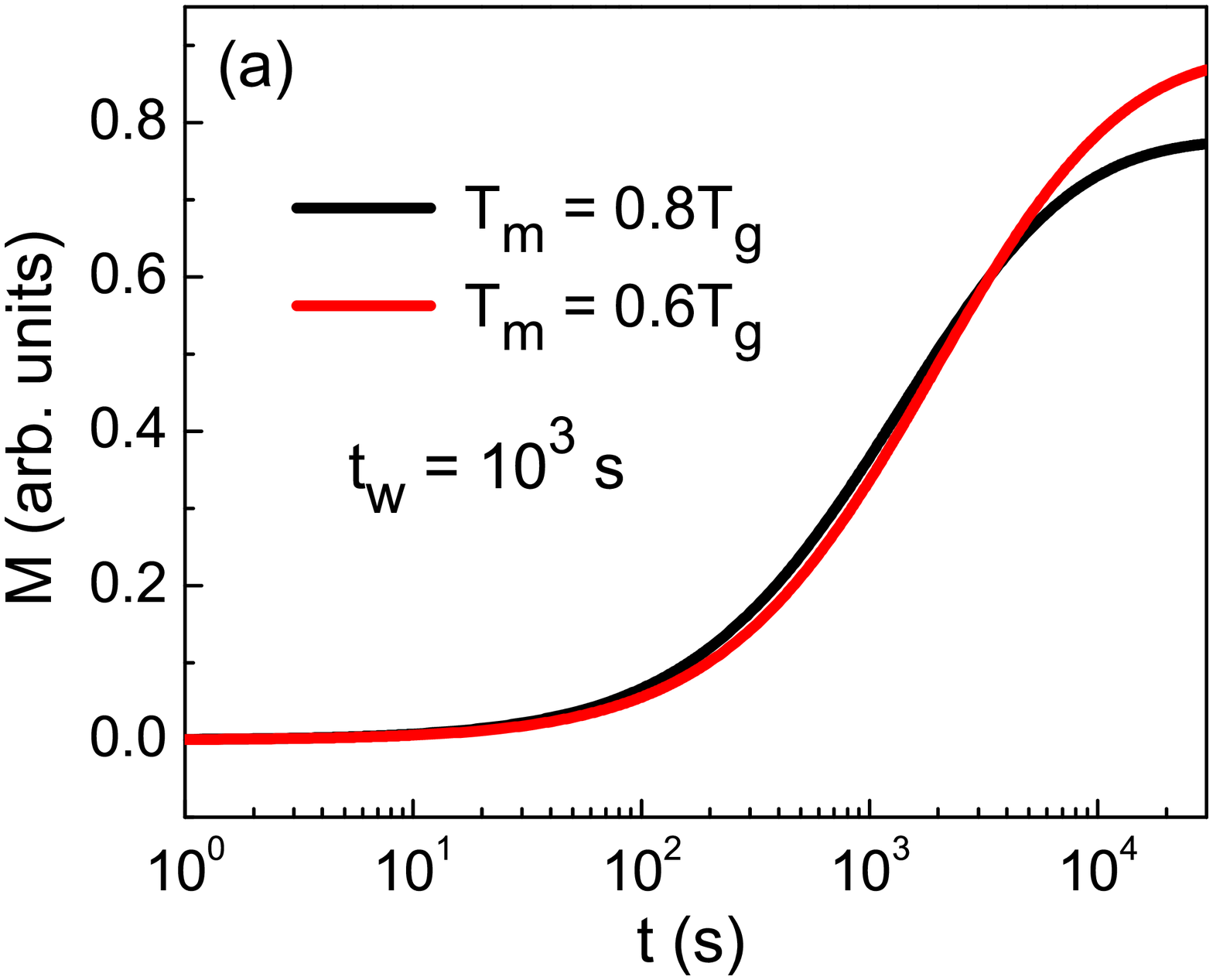}
\end{center}
\caption{(a) Comparison between two M$_{ZFC}$ curves calculated with $T_m$ = 0.8$T_g$ and 0.6$T_g$. The other parameters were kept the same at the values $A = H = c = 1$, $n$ = 0.5, $|dT/dt|$ = 0.002$T_g$, $t_w$ = 10$^{3}$ s. (b) $S$ for these M$_{ZFC}$ curves.}
\label{Fig_S4}
\end{figure}

As for the TRM case, M$_{ZFC}$ relaxation also depends on $T$, as can be seen in Eq. \ref{Eq_S3}. Fig. \ref{Fig_S4}(a) compares two curves calculated with distinct $T_m$, 0.8$T_g$ and 0.6$T_g$, but the same $A = H = c = 1$, $n$ = 0.5, $|dT/dt|$ = 0.002$T_g$, $t_w$ = 10$^{3}$ s. Fig. \ref{Fig_S4}(b) makes clear that the relaxation becomes slower as $T$ decreases, as expected.

\begin{figure}
\begin{center}
\includegraphics[width=0.6 \textwidth]{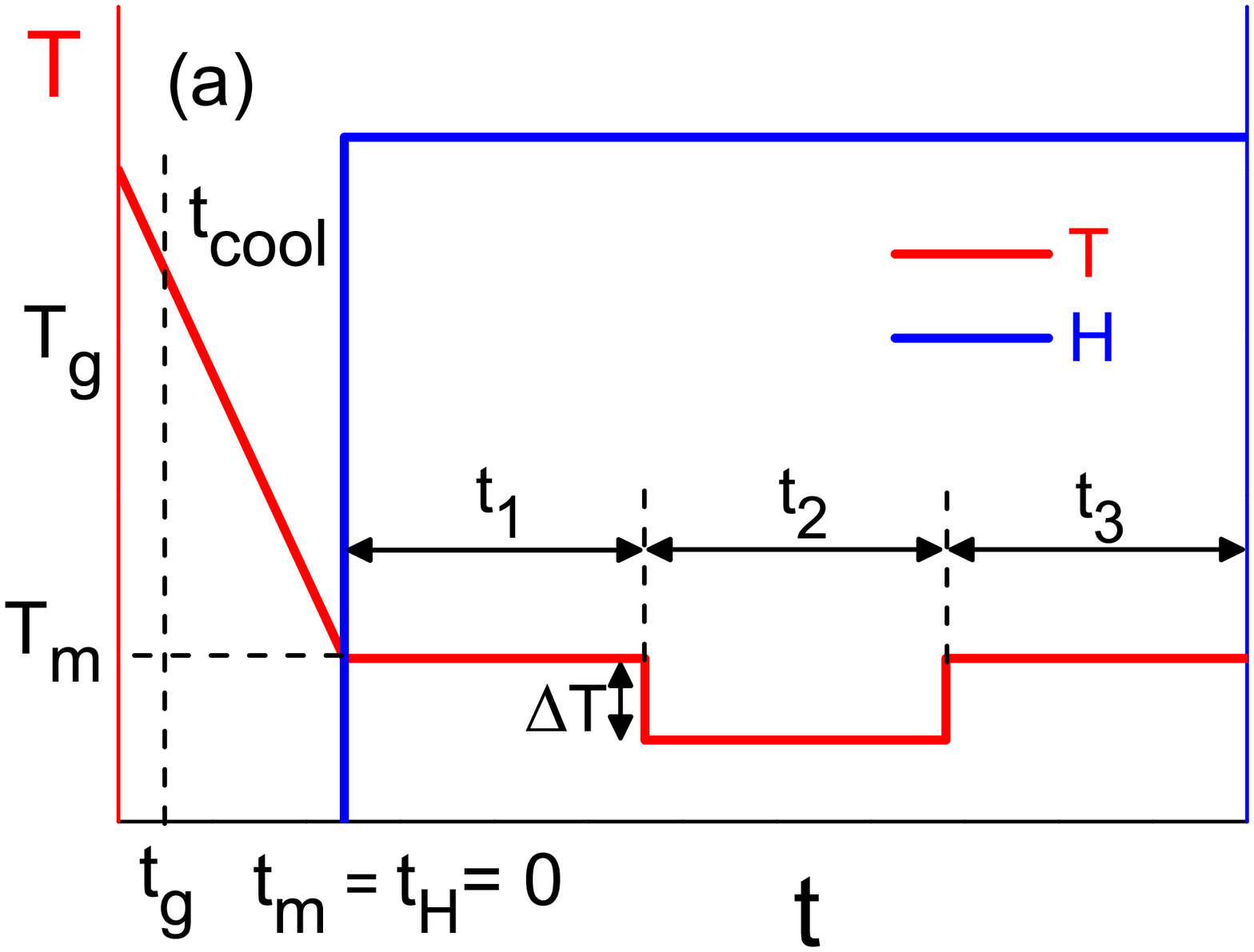}
\end{center}
\caption{(a) Protocol used to produce the $T$-cycled M$_{ZFC}$ curves where $T$ is changed during the relaxation. This figure exemplifies the $T$-cooled experiment ($\Delta T < 0$). The $T$-heated experiment is similar but with $\Delta T > 0$. (b) Resulting M$_{ZFC}$ curve when the $t_2$ stretch is removed, calculated for  $A = H = c = 1$, $n$ = 0.5, $|dT/dt|$ = 0.002$T_g$, $t_m$ = 0.5$T_g$, $\Delta T$ = -0.2$T_g$, $t_1 = t_2 = t_3$ = 4000 s. The inset shows a magnified view of the $t_1$/$t_3$ junction.}
\label{Fig_S5}
\end{figure}

For the case of the $T$-cycled M$_{ZFC}$ curves displayed in Fig. 3 of main text, the initial protocol is similar to that described above in Fig. \ref{Fig_S3} for the conventional M$_{ZFC}$ (in this case, with $t_w$ = 0). However, after a $t_1$ interval of relaxation at $T_m$, $T$ is changed to $T_m$+$\Delta T$ for a $t_2$ interval, then it returns to $T_m$ for a period $t_3$, as shown in Fig. \ref{Fig_S5}(a). The equation describing $M(t)$ under this protocol will be
\begin{equation}
\begin{split}
M(t) &= \int_{t_H = 0}^{t_1} \left( \frac{T_m}{T_g} \right)^{n}\frac{AH}{t'-t_g}e^{-\frac{cT_m}{T_g} \left( \frac{t-t'}{t'-t_g} \right)^{n}} + \int_{t_1}^{t_1+t_2} \left( \frac{T_m + \Delta T}{T_g} \right)^{n}\frac{AH}{t'-t_g}e^{-\frac{c(T_m + \Delta T)}{T_g} \left( \frac{t-t'}{t'-t_g} \right)^{n}} \\
& + \int_{t_1+t_2}^{t} \left( \frac{T_m}{T_g} \right)^{n}\frac{AH}{t'-t_g}e^{-\frac{cT_m}{T_g} \left( \frac{t-t'}{t'-t_g} \right)^{n}} \,dt'. \label{Eq_S4} \\
\end{split}
\end{equation}
Fig. \ref{Fig_S5}(b) shows a magnified view of the inset of Fig. 3(a). This is the M$_{ZFC}$ curve resulting when the $t_2$ stretch is removed, evidencing that the $t_3$ stretch seems to be a continuation of $t_1$.

\section{Magnetization as a Function of Temperature}

\begin{figure}
\begin{center}
\includegraphics[width=1.0 \textwidth]{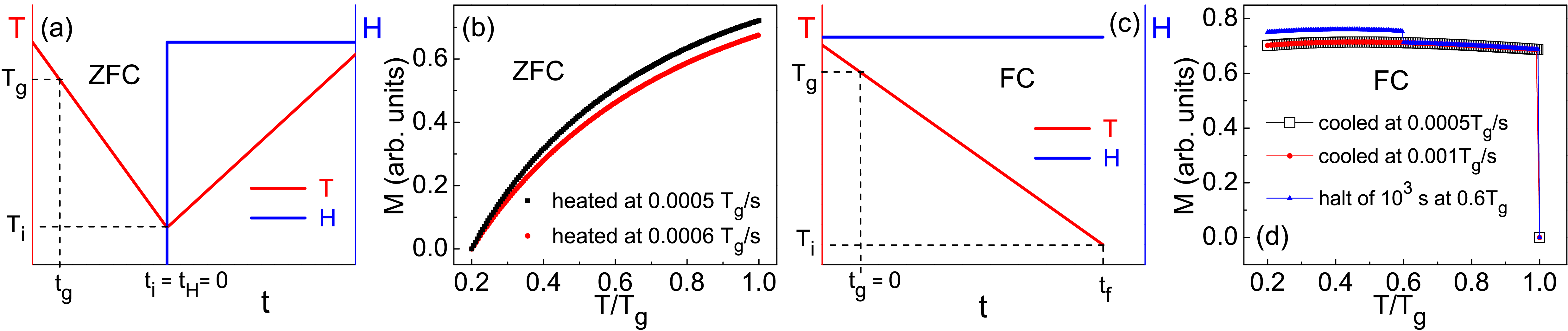}
\end{center}
\caption{(a) Protocol used to produce ZFC $dc$ M(T) curves, with $T$ decreasing and increasing in continuous mode. (b) Comparison between two ZFC M(T) curves calculated with heating $|dT/dt|$ = 0.0005$T_g$/s and 0.0006$T_g$/s, with $A = H = c = 1$, $n$ = 0.5 for both. (c) Protocol used to calculate the FC M(T) curves, with $T$ also decreasing in continuous mode. (d) Comparison between FC curves calculated with cooling $|dT/dt|$ = 0.0005$T_g$/s and 0.001$T_g$/s, $A = H = c = 1$, $n$ = 0.5. It also shows a jump in $M$ for the case in which the cooling is halted for 10$^3$ s at 0.6$T_g$.}
\label{Fig_S6}
\end{figure}

The protocol to produce the ZFC M(T) curve is the conventional one for which the system is ZFC in a constant $T$-sweep rate (for the main text it was adopted the same $|dT/dt|$ = 0.002$T_g$/s used for the TRM and M$_{ZFC}$ curves discussed above), then a dc $H$ is applied and $T$ is increased, also in a constant $|dT/dt|$, while $M$ is recorded. Fig. 4(a) of main text was calculated with $T$ increasing in the continuous mode, as shown in Fig. \ref{Fig_S6}(a). In this case, the systems' $M$ is
\begin{equation}
M(t)= \int_{t_i = 0}^{t} \frac{AH\left[ T_i+|dT/dt|(t'-t_i)\right]^n}{T_g^n(t'-t_g)}e^{-\frac{c \left[ T_i+|dT/dt|(t'-t_i)\right]}{T_g} \left( \frac{t-t'}{t'-t_g} \right)^{n}}  \,dt'. \label{Eq_S5}
\end{equation}
Since this is an off-equilibrium situation, the ZFC curve is very sensitive to changes in the $T$-sweep rate. Fig. \ref{Fig_S6}(b) shows remarkable differences between two curves calculated with slightly different heating $T$-rates, with all other parameters kept the same.

Conversely, the field cooled (FC) M(T) curves are almost invariant under changes in $|dT/dt|$. Fig. \ref{Fig_S6}(c) shows the protocol used to produce such curves, also calculated with $T$ varying in continuous mode, yielding
\begin{equation}
M(t)= \int_{t_g = 0}^{t} \frac{AH\left[ T_g-|dT/dt|(t'-t_g)\right]^n}{T_g^n(t'-t_g)}e^{-\frac{c \left[ T_g-|dT/dt|(t'-t_g)\right]}{T_g} \left( \frac{t-t'}{t'-t_g} \right)^{n}}  \,dt'. \label{Eq_S6}
\end{equation}
Fig. \ref{Fig_S6}(d) shows that the FC M(T) curves coincide even when $|dT/dt|$ is remarkably changed. However, as discussed in the main text this is not an equilibrium configuration. The figure also shows that, if the cooling is halted for a finite $t$ interval at $T < T_g$, $M$ will increase. When the cooling is resumed $M$ will maintain this increased value.

Regarding the ZFC-FC M(T) curves, there are still some import points to be addressed. The first one is that, as can be noticed from Fig. \ref{Fig_S6}, the initial $M$ value is zero for both ZFC and FC curves. This is obvious because, according to Eqs. \ref{Eq_S5} and \ref{Eq_S6}, at $t$ = 0 there was no time enough for $M$ to evolve. In practice, however, the situation is a bit different. For the ZFC case, experimentally there is an instrumental $t$ interval between the $H$ application and the initial increase of $T$. This certainly affects the systems' magnetization, leading to a non-zero $M$ value at $t_i$. For the FC curves, it must be regarded that the system is coming from a paramagnetic (PM) configuration with non-zero $M$ due to the applied $H$, which will naturally have its weight in the initial $M$ value of the SG state. Moreover, as stated in the main text, the critical behavior in the vicinity of $T_g$ will also plays its part in this region. The here proposed model is concerned with the $T < T_g$ region where the critical behavior can be neglected, so it is in fact not suitable to describe the $T$ close to $T_g$ situation. These aforementioned details, if considered here, would certainly change the slope of the M(T) curves.

It must also be noticed that Eqs. \ref{Eq_S5} and \ref{Eq_S6} give $M$ as a function of $t$. To compute the M(T) curves one must perform a change of variables, which can be easily done since $T(t) = T_i + |dT/dt|$ for the ZFC curve and $T(t) = T_i - |dT/dt|$ for the FC one.

\begin{figure}
\begin{center}
\includegraphics[width=0.6 \textwidth]{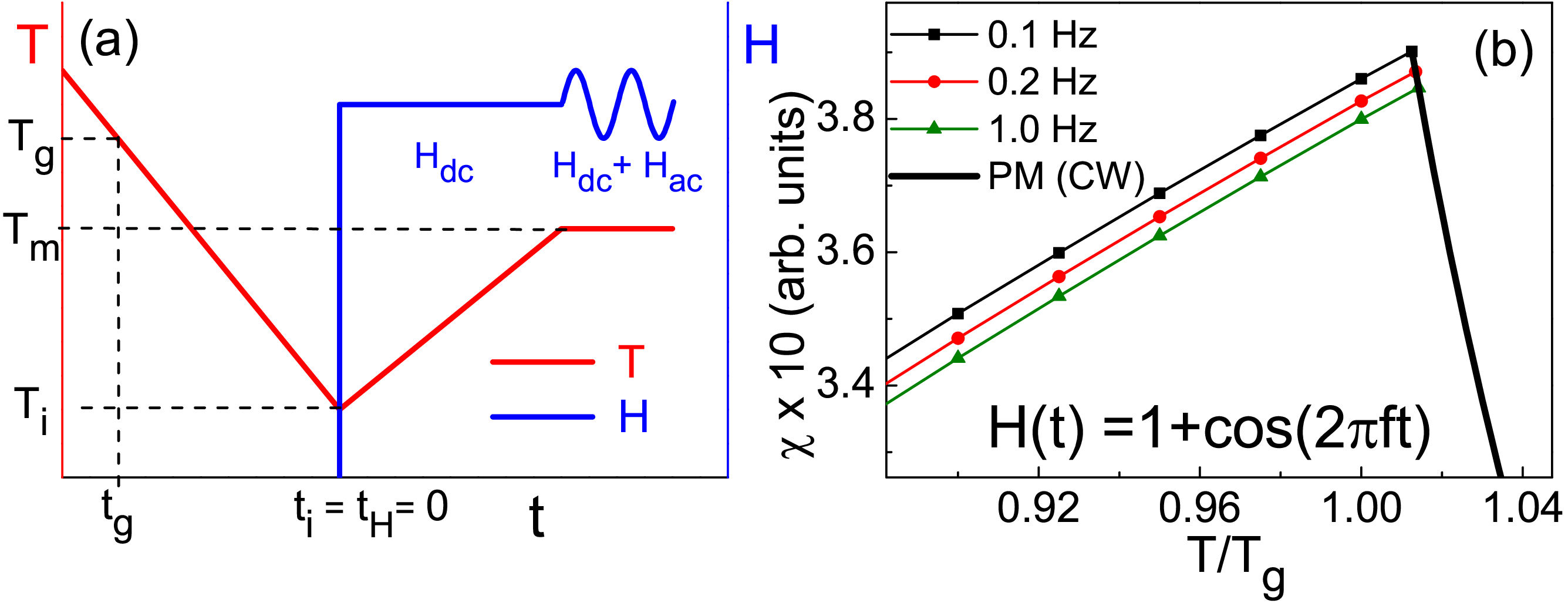}
\end{center}
\caption{(a) Protocol used to produce each point of the $ac$ M(T) curves, calculated in the heating mode. (b) $\chi_{ac}$ curves calculated with $A = H = c = 1$, $n$ = 0.5, $|dT/dt|$ = 0.002$T_g$/s and $f$ = 0.1, 0.2 and 1 Hz.}
\label{Fig_S7}
\end{figure}

The ac susceptibility ($\chi = M/H$) curves were produced point by point, \textit{i.e.} it was assumed that the system was thermalized during the measuring. The curves were calculated in the heating mode, and from one point to another $T$ was increased in a constant $dT/dt$ with the system only under the influence of a static dc field. Fig. \ref{Fig_S7}(a) shows the protocol to obtain each point, leading to the following equation
\begin{equation}
\begin{split}
M(t) & = \int_{t_i = 0}^{t_m} \frac{A \left[ T_i+|dT/dt|(t'-t_i) \right]^n }{T_g^n(t'-t_g)}e^{-\frac{ \left[ T_i+|dT/dt|(t'-t_i)\right]}{T_g} \left( \frac{t-t'}{t'-t_g} \right)^{n}}  \,dt' \\
& + \int_{tm}^{t_m+1/f} \frac{A}{t'-t_g} \left(\frac{T_m}{T_g} \right)^{n} \left\{ 1+cos[2\pi f(t'-t_m)] \right\} e^{-\left(\frac{T_m}{T_g} \right) |1+cos[2\pi f(t'-t_m)]| \left( \frac{t-t'}{t'-t_g} \right)^{n}}  \,dt'.
\label{Eq_S7} \\
\end{split}
\end{equation}
It can be noticed that it was assumed in Eq. \ref{Eq_S7} that $c = |H|$, since it was commented in main text that the stretched exponential term is expected to depend on $H$. However, very similar results are observed for the ac curves in the case that a constant $c$ value is adopted. Fig. \ref{Fig_S7}(b) shows the curves obtained for $c$ = 1, for which Eq. \ref{Eq_S7} can be adjusted to give
\begin{equation}
\begin{split}
M(t) & = \int_{t_i = 0}^{t_m} \frac{A \left[ T_i+|dT/dt|(t'-t_i) \right]^n }{T_g^n(t'-t_g)}e^{-\frac{ \left[ T_i+|dT/dt|(t'-t_i)\right]}{T_g} \left( \frac{t-t'}{t'-t_g} \right)^{n}}  \,dt' \\
& + \int_{tm}^{t_m+1/f} \frac{A}{t'-t_g} \left(\frac{T_m}{T_g} \right)^{n} \left\{ 1+cos[2\pi f(t'-t_m)] \right\} e^{-\left(\frac{T_m}{T_g} \right) \left( \frac{t-t'}{t'-t_g} \right)^{n}}  \,dt'. 
\label{Eq_S8} \\
\end{split}
\end{equation}

Again, the frequencies ($f$) were chosen slow enough so that one can assume $\chi = M/H$ as a good approximation. The relative shift $\delta T_f$ = $\Delta T_f/T_f(\Delta log f)$ obtained for the curves of Fig. \ref{Fig_S7}(b) is $\sim$ 0.002, similar to the value found with $c$ = $|H|$. I must recall that the here proposed model does not predict the PM-SG transition, the $T_f$ was here assumed as the point in which the SG and PM curves intercept, these values being thus closely related to the choice of the PM curve. Usually the $T > T_g$ curve is flattened in relation to the CW law in the vicinity of $T_g$, which would lead to a larger $\delta T_f$.

\section{Different Functional Forms for $b$ and $M_0$}

\begin{figure}[h]
\begin{center}
\includegraphics[width=1.0 \textwidth]{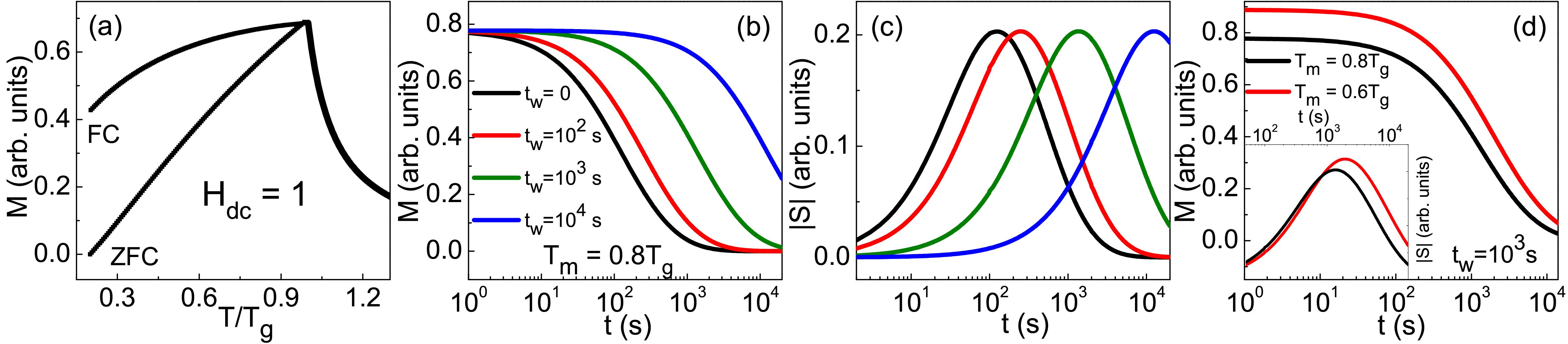}
\end{center}
\caption{Some curves calculated with Eq. \ref{Eq_S9}, using $A = H = c = 1$, $n$ = 0.5: (a) ZFC-FC M(T) computed with cooling $|dT/dt|$ = 0.001$T_g$/s. For the ZFC curve it was also used a heating $|dT/dt|$ = 0.0006$T_g$/s. (b) TRM curves calculated for different $t_w$ at $T_m$ = 0.8$T_g$. (c) The $|S|$ of the TRM curves for different $t_w$. (d) Comparison between TRM curves at $T_m$ = 0.8$T_g$ and 0.6$T_g$, calculated with $t_w$ = 10$^{3}$ s. The inset shows $|S|$ for these curves.}
\label{Fig_S8}
\end{figure}

Eqs. 1 and 2 of main text gives the systems' $M$ at an instant $t$ due to a $H$ that was applied at a previous instant $t$'. It may occur the situation in which $T$ at the instant that $H$ was applied is different from that at instant $t$, \textit{i.e.} $T(t') \neq T(t)$. This will surely affect the magnetization since changes in $T$ alter the free energy landscape, thus affecting the systems' position in this landscape and the $M$-decay. A natural step here is to consider the possibility of $T$ being a function of $t$ instead of $t'$ in the equations for $M_0$ and $b$. Lets consider first the case in which $T(t')$ is replaced by $T(t)$ in Eq. 2 for $M_0$, yielding
\begin{equation}
M_0 = \left[\dfrac{T(t)}{T_{g}}\right]^n\frac{AH}{t'-t_{g}}. \label{Eq_S9}
\end{equation}
Eqs. \ref{Eq_S5} and \ref{Eq_S6} can be adapted to give the ZFC-FC M(T) curves for this case, displayed in Fig. \ref{Fig_S8}(a). Interestingly, the ZFC is similar to that found using $T(t')$ in $M_0$ whereas the FC one decreases with $T$. The fundamental difference between these two approaches is that with Eq. \ref{Eq_S9} we are considering that, although $H$ was applied at $t'$, the systems' $M$ is immediately related to the $M_0$ value at $t$ [and consequently to the $T(t)$ value]. Conversely, with Eq. 2 of main text we compute $M(t)$ as a consequence of the $M_0(t')$ value. In practice, in a dynamic situation like that of M(T) measurements the systems' decay may occur continuously between $t'$ and $t$ [\textit{i.e.} between $T(t')$ and $T(t)$], and the resulting curve may lies between those computed with $T(t')$ and $T(t)$.

For the ac curves, in this work each point was computed after the system being thermally stabilized. Thus one can expect the same overall behavior independently of using Eq. 2 of main text or Eq. \ref{Eq_S9}. For the calculation of TRM, it is interesting to note that, in spite of the non-negligible changes in the equation for the $t_{cool}$ interval when $T(t')$ is replaced by $T(t)$ in $M_0$, the resulting curves displayed in Fig. \ref{Fig_S8}(b)-(d) present the same features of those observed in the main text. Finally, for the M$_{ZFC}$ experiment, since in this case $H$ is applied when the system is already stabilized at $T_m$, the resulting curves will be precisely the same as those displayed in Fig. 2 of main text.

\begin{figure}[H]
\begin{center}
\includegraphics[width=1.0 \textwidth]{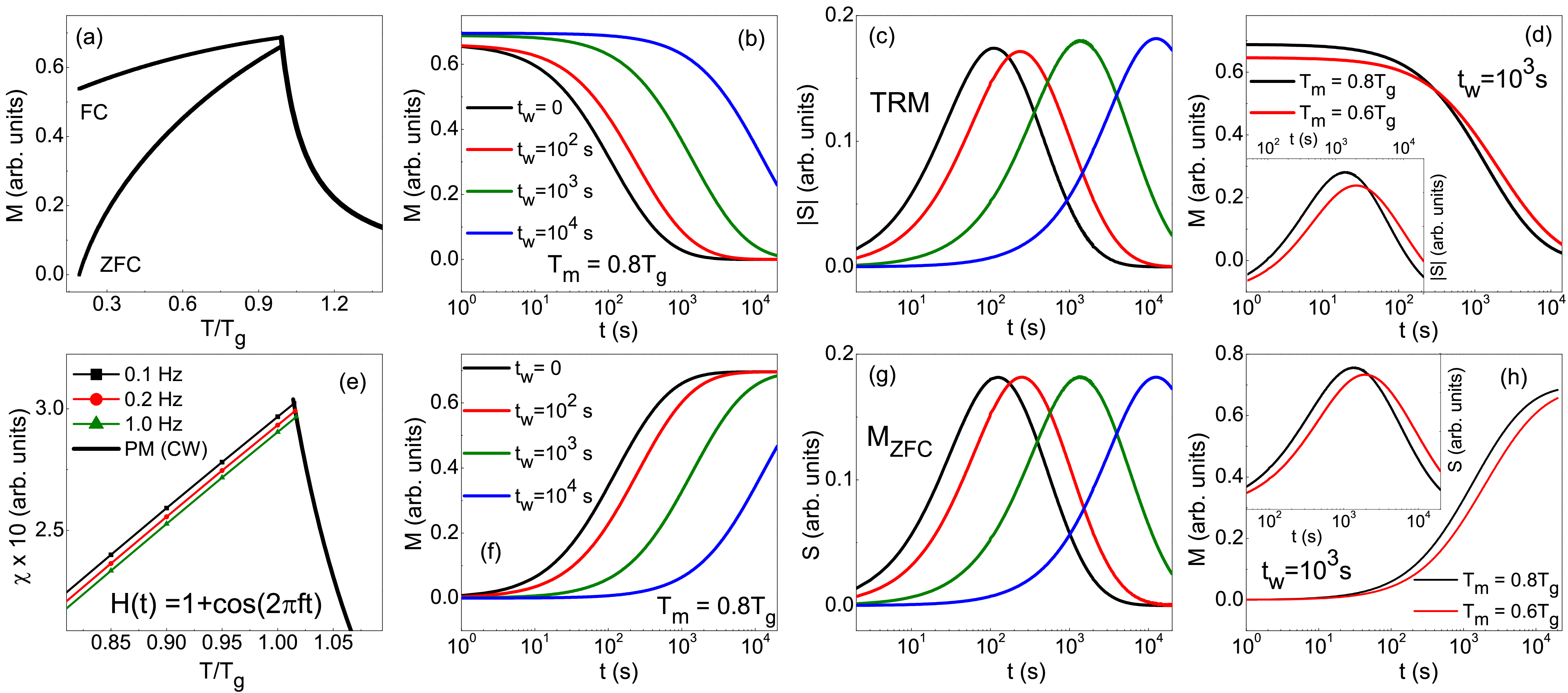}
\end{center}
\caption{Main results obtained with Eq. \ref{Eq_S10}, using $A = H = c = 1$, $n$ = 0.5: (a) ZFC-FC M(T) computed with cooling $|dT/dt|$ = 0.002$T_g$/s. For the ZFC curve it was also used a heating $|dT/dt|$ = 0.0002$T_g$/s. (b) TRM curves calculated for different $t_w$ at $T_m$ = 0.8$T_g$. (c) The $|S|$ for these TRM curves with different $t_w$. (d) Comparison between TRM curves at $T_m$ = 0.8$T_g$ and 0.6$T_g$, calculated with $t_w$ = 10$^{3}$ s. The inset shows $|S|$ for these curves. (e) $\chi$ $ac$ curves calculated for $f$ = 0.1, 0.2 and 1 Hz. (f) M$_{ZFC}$ curves calculated for different $t_w$ at $T_m$ = 0.8$T_g$, and (g) the $S$ of these curves. (h) Comparison between M$_{ZFC}$ curves at $T_m$ = 0.8$T_g$ and 0.6$T_g$, calculated with $t_w$ = 10$^{3}$ s. The inset shows $S$ for each $T_m$.}
\label{Fig_S9}
\end{figure}

These results indicate that the most important finding of this paper is the stretched exponential Eq. 1 of main text, which must be integrated along the interval at which the system is under the influence of $H$, whereas the equations for $M_0$ and $b$ can in principle be adapted to better describe each material. Another possible functional form for $M_0$, for instance, is that one for which one removes the $n$ exponent in the $T(t')/T_g$ term of Eq. 2, yielding
\begin{equation}
M_0 = \left[\dfrac{T(t)}{T_{g}}\right]\frac{AH}{t'-t_{g}}. \label{Eq_S10}
\end{equation}
Fig. \ref{Fig_S9} displays the main results obtained using Eq. \ref{Eq_S10}. The FC M(T) curve is similar to that obtained with Eq. \ref{Eq_S9},and the evolution of the TRM curves with $t_w$ is also very similar to those observed when Eqs. 2 and \ref{Eq_S9} are used, although the decrease in $M$ is less pronounced here. But an interesting difference between the results of Eq. \ref{Eq_S10} and those obtained with the previous functions, shown in Fig. \ref{Fig_S9}(d), is that the TRM curve calculated for $T_m$ = 0.6$T_g$ starts below the $T_m$ = 0.8$T_g$ one, as experimentally observed for some materials. Since the decay is faster for the 0.8$T_g$ curve (as expected), the curves will intercept at some point. The ac $\chi$ curves present the expected $f$-dependence, and the results for $M_{ZFC}$ also show the expected overall behavior.

\begin{figure}[H]
\begin{center}
\includegraphics[width=1.0 \textwidth]{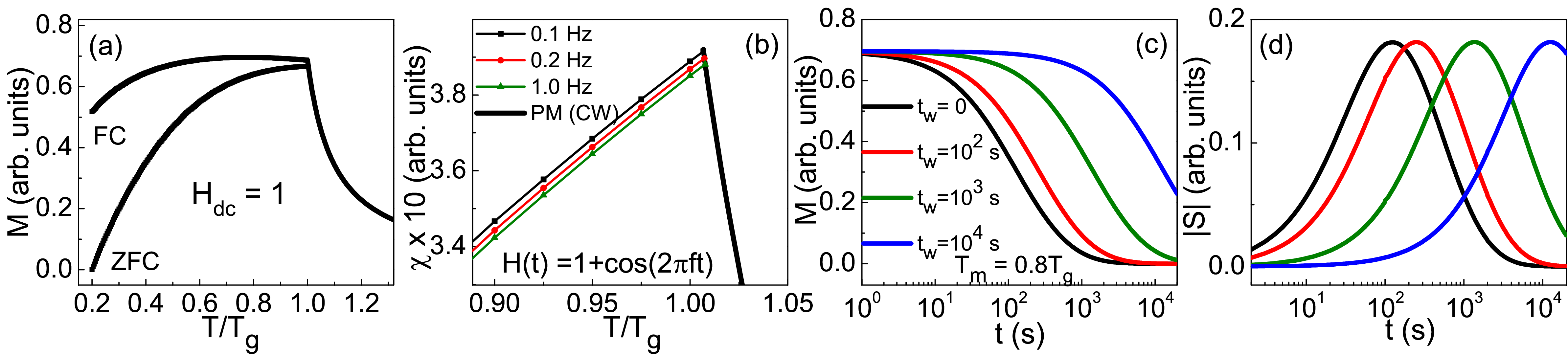}
\end{center}
\caption{Main results obtained with Eq. \ref{Eq_S11}, using $A = H = c = 1$, $n$ = 0.5: (a) ZFC-FC M(T) curves computed with cooling $|dT/dt|$ = 0.002$T_g$/s. For the ZFC curve it was also used a heating $|dT/dt|$ = 0.0002$T_g$/s. (b) $\chi$ $ac$ curves calculated for $f$ = 0.1, 0.2 and 1 Hz. (c) TRM curves calculated for different $t_w$ at $T_m$ = 0.8$T_g$. (c) $|S|$ of the TRM curves calculated with different $t_w$.}
\label{Fig_S10}
\end{figure}

One can also replace $T(t')$ by $T(t)$ in both $M_0$ and $b$, leading for instance to 
\begin{equation}
M_0 = \left[\dfrac{T(t)}{T_{g}}\right] \frac{AH}{t'-t_{g}}; b = \frac{c \cdot T(t)}{T_{g}(t'-t_{g})^n}.  \label{Eq_S11}
\end{equation}
Figs. \ref{Fig_S10}(a) and (b) displays respectively the dc and ac M(T) curves and the main TRM results obtained when Eqs. \ref{Eq_S11} are used in Eq. 1. The dc FC M(T) curve shows a plateau-like behavior followed by a small decrease in $M$ with decreasing $T$, while the ac $\chi$ curves show the expected $f$-dependency. The TRM curves are also very similar to those observed for the other functions here discussed, with a maxima in $|S|$ around $t_w$, and since for the M$_{ZFC}$ curves $H$ is applied when the system is already stabilized at $T_m$ the resulting curves will be precisely the same as those obtained with Eq. \ref{Eq_S10}, Fig. \ref{Fig_S9}.

Surely, each function here discussed must be carefully investigated in order to check if it is in fact suitable to describe SG-like systems and to ensure that it is scientifically sound. Nevertheless, the agreement between the results here obtained and the main experiments reported along these almost five decades of investigation is remarkable.